\documentclass[useAMS,usenatbib,usegraphicx]{mn2e}

\newcommand{\ergps}{erg~s$^{-1}$}

\newcommand{\etal}{et al.}
\newcommand{\mcg}{MCG--6-30-15}
\newcommand{\h}{1H\,0419--577}

\title[X-ray reflection in the Seyfert galaxy 1H\,0419-577]
  {X-ray reflection in the Seyfert galaxy 1H\,0419-577 revealing
    strong relativistic effects in the vicinity of a Kerr black hole}
\author[A.\ C.\ Fabian \etal]
  {A.~C.~Fabian$^1$\thanks{acf@ast.cam.ac.uk}, G.~Miniutti$^1$,
  K.~Iwasawa$^1$ and R.R.~Ross$^2$\\
  $^1$Institute of Astronomy, Madingley Road, Cambridge CB3 0HA \\
    $^2$Physics Department, College of the Holy Cross, Worcester, MA
  01610, USA}

\pagerange{\pageref{firstpage}--\pageref{lastpage}}
\pubyear{2001}

\usepackage{times}

\begin{document}

\label{firstpage}

 \maketitle

\begin{abstract}
  We report results obtained from six XMM--Newton observations of the
  Seyfert galaxy 1H~0419--577. The source was observed in a wide range
  of different flux levels, allowing its long--term spectral
  variability to be studied in detail as already reported by Pounds et
  al (2004a; 2004b). Here we show that the X--ray spectrum is well
  described by a simple two--component model comprising a power law
  with constant spectral shape and variable normalisation and a much
  more constant ionised reflection component from the inner accretion
  disc which carries the signature of strong relativistic effects. One
  of the observations was performed when the source was in a
  particularly low flux state in which the X--ray spectrum is rather
  peculiar and exhibits a very flat hard spectrum (with spectral index
  close to 1 in the 2--10~keV band) with broad residuals below 6.6~keV
  (rest--frame) and a steep soft excess below about 1~keV. We
  interpret the spectrum as being reflection--dominated by X--ray
  reprocessed emission from the inner accretion disc. The primary
  continuum, which illuminates the disc, is almost completely
  unobserved possibly because of strong light bending towards the
  central super--massive black hole. The ionised reflection model
  simultaneously accounts for the broad hard residuals and hard flat
  spectrum and for the soft excess.  The same model provides an
  excellent description of the data at all the other flux levels, the
  most important difference being a variation in the power law
  normalisation. Our spectral decomposition and interpretation of the
  spectral variability implies that most of the X--ray emission in
  this source originates from within few gravitational radii from the
  central black hole and requires that the compact object is almost
  maximally spinning.
\end{abstract}


\section{Introduction}
\label{sect:intro}

\h\ is an EUV-luminous, radio-quiet, optically broad-line, Seyfert
1--1.5 galaxy at redshift $z=0.104$ (Pye et al 1995; Marshall, Fruscione
\& Carone 1995). Its 2--10~keV luminosity of $5\times 10^{44}$ \ergps
(Guainazzi et al 1998; Page et al 2002) puts it around the
Seyfert--quasar borderline. It shows remarkable long-term soft X-ray
variability ranging from a very steep to a relatively flat spectrum
(Guainazzi et al 1998; Turner et al 1999; Page et al 2002). 

The source has been observed several times with XMM-Newton, and
previous analysis by Pounds et al (2004a; 2004b) revealed that the
dominant spectral variability is due to a steep power-law component. 
\h\ has a high state dominated by this power law, of photon index
$\Gamma \simeq 1.8$, and a low state where the spectrum is unusually
very flat, $\Gamma\sim 1$ in the 2--10~keV band. Pounds et al obtain
good fits to the low-state spectrum with either a very broad iron-K
line and continuum reflection or partial covering. A strong blackbody
component was also included to model a prominent soft excess below
about 1~keV. For the reflection fit, a very high reflection factor,
$R\sim 3.5$, was required and the irradiating power-law is still very
flat $\Gamma\sim 1.26$). Guainazzi et al (1998) found what they
considered to be an improbably high reflection fraction ($R\sim 10$)
from their spectral fitting of BeppoSAX data. ASCA data revealed for
the first time the presence of a Fe line in \h\ (Turner et al 1999). 

The possibility of a reflection--dominated spectrum in the low flux
state of \h\ matches very well the predictions of a light-bending
model we recently developed for the strong gravity regime in the
well-studied Seyfert galaxy \mcg\ (Fabian \& Vaughan 2003; Miniutti et
al 2003; Miniutti \& Fabian 2004; Vaughan \& Fabian 2004) and for the
very high state of the Galactic black hole Candidate XTE~J1650--500
(Miniutti, Fabian \& Miller 2004; Rossi et al. 2005). This model
consists of power--law (PLC) and reflection-dominated (RDC) components
and was able also to explain the complex spactrum and variability of
the NLS1 1H\,0707-495 (Fabian et al. 2002; 2004). The PLC is assumed
to originate in a compact region at a few gravitational radii ($r_{\rm
  g}=GM/c^2$) above the innermost disc. The emitting region can be
either static or comoving with the accretion flow, point--like or
ring--like, withouth much differences in the resulting behaviour. In
Miniutti \& Fabian (2004) we assumed the PLC originates from a
ring--like structure at $2~r_g$ from a Kerr black hole axis, but
similar results are obtained for compact sources within about $4~r_g$
and even for a primary source on the rotation axis of the black hole
(such as in the lamp--post model, see e.g. Martocchia, Karas \& Matt 2000).

The ring--like configuration is not necessarly the actual emitting
structure we have in mind at any given time, but the time--averaged
configuration over the timescale needed to extract good quality
spectra from present X--ray observatories. A collection of
static/corotating sources related e.g. to magnetic dissipation
processes preferentially occurring at few $r_g$ from the hole in a
patchy corona, X--ray emission in the base of a compact jet (e.g.
Markoff, Falcke \& Fender 2001), or due to internal shocks from
aborted jets close to the hole axis (Ghisellini, Harrdt, Matt 2004)
might all be viable physical mechanisms, the list being far from
exhaustive.

We note here that evidence for an orbiting reflecting spot, possibly
related to a corotating illuminating flare above it (one of the
primary sources?) at about $10~r_g$ from the black
hole in NGC~3516 has been reported recently (Iwasawa, Miniutti \&
Fabian 2004). Flares/emitting regions located even closer to the
central mass, would prove difficult to follow in reflection with
current detectors, especially if lasting, as it is likely due to
turbulence and shear in the innermost flow, less than one orbital
period. A reasonable zeroth--order approximation for the
time--averaged emitting source geometry seems that of assuming a ring
representing globally the region of preferential energy dissipation
since the natural symmetry is around the black hole axis.

The main idea of the light bending model is that, since the primary
source is in a region of strong gravity, the flux of the observed PLC
is reduced by gravitational light bending which focus radiation down
onto the disc where the RDC originates (the closer the source to the
black hole, the stronger the effect). The PLC can then appear to vary
in amplitude by a large factor (an order of magnitude or more) as the
location of the primary source is allowed to vary, even its intrinsic
luminosity is constant, while the RDC changes only little. The
spectrum of the RDC is affected by the redshifts expected from a
component situated only a few $r_{\rm g}$ from a black hole. The main
prediction of this model is that the spectrum becomes more and more
reflection--dominated as the PLC flux drops.

Here we apply the light-bending model to the XMM-Newton data of \h. We
use new ionised slab models for the reflection component (Ross \&
Fabian 2005) based on previous calculations by Ross et al (1993;
1999; 2002), the whole spectrum of which is blurred relativistically
to include the effects of emission originating in a geometrically
thin accretion disc around a rotating black hole. Our aim is to see
whether a simple two-component (PLC plus RDC) light-bending model
explains the whole broadband spectrum of \h\ and its variability. 

\section{The XMM--Newton observations}

XMM--Newton observed \h\ six times between December 2000 and September
2003 (orbits 181, 512, 558, 605, 649, and 690). The data have been
kindly provided to us by K.A. Pounds (see Pounds et al 2004b for
details on data reduction) except for orbit 181 whose data have been
downloaded from the XMM--Newton Science Archive. We only make use here
of the EPIC--pn spectra which were grouped to a minimum of 20 counts
per energy bin to allow the use of the $\chi^2$ minimisation during
spectral fitting. Errors are quoted at the 90 per cent confidence
level for one interesting parameter.

\section{The low flux state of \h: XMM--Newton orbit 512}

As reported by Pounds et al (2004a) one of the XMM--Newton
observations (orbit 512) caught \h\ in a low flux state. Low flux
states are of great interest in the framework of the light bending
model because they are predicted to be dominated by strong disc
reflection from the accretion disc. This is because low flux states
correspond to geometries in which the primary X--ray source is very
close to the black hole, so that most of its emission is bent onto the
accretion disc, reducing the continuum flux that reaches the observer
at infinity directly while at the same time enhancing the disc
illumination, and therefore the reflection fraction (Miniutti \&
Fabian 2004). 

When a simple power--law model is fitted to the data in the 2--10~keV
band, a very flat photon index ($\Gamma \sim 1$) is obtained leaving
clear broad residuals below about 6~keV in the observer frame,
corresponding to about 6.6~keV in the rest--frame. Re--inserting the
0.5--2~keV data reveals a strong soft excess below about 1~keV. The
residuals (in terms of $\sigma$) of such a fit are shown in Fig.~1
where the soft excess and broad hard residuals are seen. 

As reported by Pounds et al (2004a) the soft excess can be described
by black body emission, while the broad hard residuals can be
accounted for with a partial covering model by slightly ionised
matter.  In such a partial covering fit, narrow iron emission is also
detected at $E = 6.21 \pm 0.11$ in the source rest--frame, barely
consistent with the iron K$\alpha$ energy ($\ge$~6.4~keV). The
observed redshifted energy is at odds with the partial covering
interpretation given that the absorbing column should produce neutral
or even slightly ionised iron emission. The iron line energy could
still arise from the absorbing matter if the absorber originates in an
outflow (but where is the blueshifted component?), in an inflow, or if
it is located close to the black hole and suffers gravitational
redshift.  However, the absorption edge is not redshifted by the same
amount as the line, but lies at $7.10\pm 0.06$~keV in the source
rest--frame giving rise to a somewhat inconsistent picture (though at
the 90 per cent level only). 

Pounds et al (2004a) also explored a different interpretation of the hard
spectrum in terms of broad relativistic iron emission and strong
reflection which gave a comparable good fit to the data as the partial
covering model with a large line equivalent width of about 1~keV. The
power--law photon index was still very flat as with the partial
covering model, challenging Comptonisation models for the X--ray
continuum production (Haardt \& Maraschi 1991).  However, the use of
an inconsistent reflection model (relativistic line at $6.9$~keV plus
a neutral reflection continuum which is not relativistically blurred)
might have affected the spectral results. 
\begin{figure}
  \includegraphics[width=0.30\textwidth,height=0.46\textwidth,angle=-90]{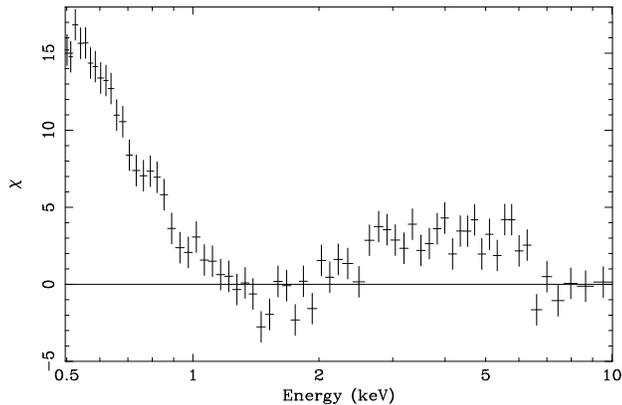}
  \caption{Ratio (in terms of $\sigma$) of the data of orbit 512 to a
    power law model fitted in the 2--10~keV band only. The figure
    shows the presence of a soft excess below about 1~keV and broad
    residuals between 3 and 6~keV (observer frame, $z=0.104$).  }
\end{figure}

Here we apply a spectral decomposition inspired by the light bending
model to the spectrum of the low flux state. We consider the
0.5--10~keV data, because of some residual calibration uncertainties
in the EPIC--pn soft energy band (Kirsch et al 2004).  The model
comprises only two broadband components, namely a power-law, and an
ionised reflection model where reflection continuum and emission lines
are computed self--consistently (Ross \& Fabian 2005). The reflection
spectrum is the result of a power--law illumination of Compton thick
material, and we tied its photon index to that of the visible
power--law component.  The reflection model allows for variable iron
abundance that was left as a free parameter of the model. The
relativistic blurring makes use of the {\tt LAOR} kernel describing
relativistic effects on the spectral shape resulting from emission in
an accretion disc orbiting a Kerr black hole (Laor 1991). The outer
disc radius ($r_{\rm{out}}$) was fixed at its maximum possible value
of $400~r_g$, while the inner disc radius ($r_{\rm{in}}$) and disc
inclination were allowed to vary. The emissivity profile, describing
the dependency of the emissivity $\epsilon$ with the radial position
on the disc, has the form of a broken power law with $\epsilon =
r^{-q_{\rm{in}}}$ for $r_{\rm{in}} \leq r \leq r_{\rm{br}}$ and
$\epsilon = r^{-q_{\rm{out}}}$ for $r_{\rm{br}} \leq r \leq
r_{\rm{out}}$, where $r_{\rm{br}}$ is a break radius on the disc. 

All spectral fits include photoelectric absorption fixed at the
Galactic value ($N_{\rm H} = 2\times 10^{20}$~cm$^{-2}$) and an
additional absorption component at the redshift of \h\ to account for
possible excess absorption local to the source. Our aim is to explore
whether the ionised reflection model can simultaneously account for
the hard broad residuals, the flat spectrum, and the soft
excess below 1~keV and to see whether a more consistent reflection
model can remove the requirement for the implausibly flat 2--10~keV 
spectral shape. 

\begin{figure}
\includegraphics[width=0.32\textwidth,height=0.46\textwidth,angle=-90]{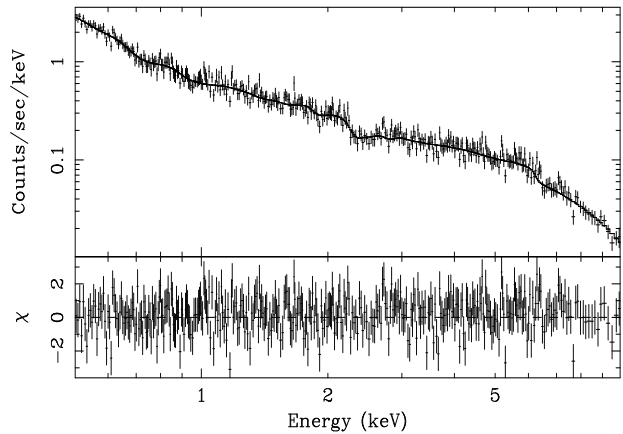}
\caption{The 0.5-10~keV spectrum and residuals (in terms of $\sigma$)
  for the RDC plus PLC model of the LF state data (orbit 512),
  producing an excellent fit with $\chi^2 = 626$ for 638 degrees of
  freedom.  }
\end{figure}

\begin{figure}
\includegraphics[width=0.30\textwidth,height=0.46\textwidth,angle=-90]{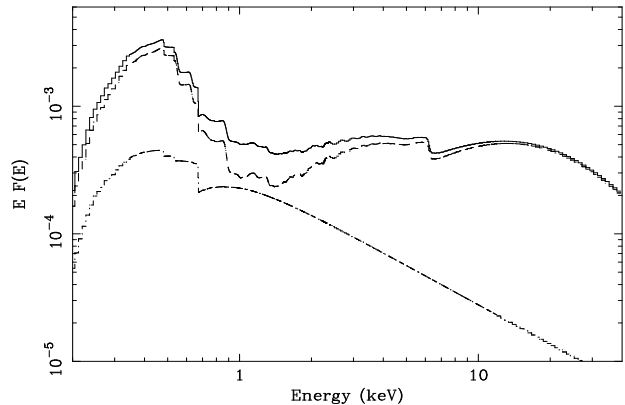}
\caption{The best--fit model for the LF state of orbit 512 (see
  Fig.~2) is shown in the 0.2--40~keV band. The two spectral
  components are shown, with the RDC dominating both the soft and hard
  bands, with some PLC contamination in the intermediate band. The sum of
  the two spectral components, i.e. the total best--fit model, is also
  shown. The resulting reflection--dominated spectrum explains
  simultaneously both the flat spectral shape and broad residuals in
  the hard band and the soft excess below 1~keV.  }
\end{figure}

With the addition of an absorption edge at $\sim 0.74$~keV, consistent
with O {\footnotesize VII} in a warm absorber, our model provides a
very good description of the low flux state data with $\chi^2=626$ for
638 degrees of freedom and null hypothesis probability of 0.636. No
additional edge at $7.1$~keV is required nor an absorption component
above 2~keV. Both the soft excess and the hard spectrum are very well
described by the relativistically blurred ionised reflection model.
The 0.5--10~keV spectrum and best-fit residuals (in terms of $\sigma$)
are shown in Fig.~2. The best--fit power law has a photon index of
$\Gamma \simeq 2.2$, much steeper than previously derived (Pounds et
al. 2004a) and totally consistent with standard Comptonisation models.
As mentioned, this is most likely the effect of using a
self--consistent reflection model in which emission lines and the
reflection continuum are computed together and affected by the the
same Doppler and gravitational shifts in the accretion disc.

In Fig.~3, we show the best--fit model in the (broader than fitted)
energy range between 0.2~keV and 40~keV.  From the Figure, the origin
of the hard (2--10~keV) flat spectrum and steep soft excess below
1~keV is clear and due to the RDC. The spectrum is almost completely
reflection--dominated with only little contribution from the
power--law. This means that the reflection fraction
(i.e. the relative amplitude of the RDC with respect to the PLC
continuum) is much larger than unity and that the reflector is seeing
much more illuminating flux than we detect at infinity as the PLC of
the spectrum. This is expected if the primary PLC emission is strongly
anisotropic and preferentially shining towards the disc. One possibility is provided by strong
light bending which is effective if the primary PLC source is located
only few $r_g$ from the black hole producing low flux and
reflection--dominated states. In this case, the RDC is
emitted from the inner regions of the disc and is therefore
strongly distorted by relativistic effects. 

The parameters of the relativistic blurring support an interpretation
in terms of strong relativistic effects.  The emissivity profile
($\epsilon=r^{-q}$) is best described by a broken power--law with
$q_{\rm{in}}=7.0 \pm 0.7$ within the break radius
$r_{\rm{br}}=3.5^{+2.3}_{-0.7}~r_g$ and $q_{\rm{out}}=3.7 \pm 0.6$ at
larger radii. The steep emissivity indicates that the disc is
illuminated mostly in its very inner regions supporting the idea that
the primary source is very close to the black hole. This is the region
where light bending effects are most effective (and actually
unavoidable) in reducing the PLC at infinity and therefore in
producing a reflection--dominated spectrum.  We also point out that if
this is the case, the emissivity profile would indeed result to be
well approximated by a broken power law, much steeper in the inner
disc and flatter in the outer regions (Martocchia, Karas \& Matt 2000;
Miniutti \& Fabian 2004), as observed in this low flux state of \h. 

The observer inclination is not very well constrained (being
degenerate with the line energy or disc ionisation state) with a
best--fit value of $36\pm 10$ degrees. The best--fit inner disc radius
is $1.4~r_g$ with a 90 per cent upper limit at $2~r_g$, very close to
the innermost stable circular orbit around a maximally rotating Kerr
black hole ($\sim 1.24~r_g$), strongly suggesting that the black hole
in \h\ is rapidly rotating. The error on $r_{\rm{in}}$ is so small
because emission from the innermost regions of the accretion disc is
necessary to reproduce the broad curvature of the spectrum in the Fe K
band.

The reflection model requires iron to be overabundant with respect to
solar, the best--fit value being $(~3.8\pm 1.5~)\times$~solar.
Together with the large reflection fraction, this explains the large
value of the equivalent width of the broad iron line (about 1~keV)
obtained by Pounds et al (2004a) when fitting the hard residuals with
a relativistic line profile. The high iron abundance we find is not
unique to this object and is required e.g. in MCG--6-30-15 and
1H~0707-495 (Vaughan \& Fabian 2004; Fabian et al 2002; 2004), also
well described by a similar spectral model. We note that Shemmer et al
(2004) recently found a metallicity--accretion rate correlation by
studying a sample of active galaxies in the near--infrared (see the
reference for possible interpretations). The mass of the black hole in
\h\ can be estimated by using the Kaspi et al (2000) relationship
between H$\beta$ line width, $\lambda L_{\lambda}$ (5100 \AA)
luminosity, and black hole mass. The optical measurements for \h\ are
reported in Grupe et al (2004) and we estimate a black hole mass of
$1$--$2 \times 10^{8}~M_\odot$ (see also Pounds et al 2004b), which
gives $L_{\rm{Edd}} = 1.3$--$2.6 \times 10^{46}$~erg~s$^{-1}$. By
considering the bolometric luminosity of $2.4\times
10^{46}$~erg~s$^{-1}$ (Grupe et al 2004), \h\ might indeed be
accreting at very high rate.  It is therefore possible that
MCG--6-30-15, 1H~0707-495, and \h\ are all accreting at high rate and
mainly differ by a different black hole mass, much higher in \h\ as
the lack of rapid large amplitude X--ray variability (as opposed to
the other two sources) also indicates.

Finally, from the reflection model we measure an ionisation parameter
of $\xi = 45 \pm 18$~erg~cm~s$^{-1}$ for the reflector surface. However, since the
radial dependency of the emissivity profile implies that the inner
disc is much more illuminated than the outer regions, it is very
likely that the ionisation structure on the disc surface is also
strongly dependent on $r$. The most naive expectation (for a disc with
approximately constant density) is that the disc is more ionised in
the inner than outer regions, while we have used so far a reflection
model with uniform ionisation parameter.  We then replaced our simple
reflection model by a composite one, in which the inner disc is
allowed to have a different ionisation state than the outer one, with
a break radius $r_{\rm{br}}$ where the ionisation parameter and the
emissivity profile change abruptly. This is of course a zeroth--order
approximation for a real situation in which the ionisation parameter
is a function of $r$. We obtain a very good description of the data
with $\chi^2 = 620$ for 637 degrees of freedom, which represents a
modest but somewhat significant improvement on the single--reflector
fit (98.7 per cent according to the F--test). The best--fit ionisation
parameters are $\xi_{\rm{in}} = 69 \pm 20$~erg~cm~s$^{-1}$ in the inner disc from
$1.24~r_g$ to $4.7~r_g$, and $\xi_{\rm{out}} = 29 \pm 16$~erg~cm~s$^{-1}$ out to
$400~r_g$. As in the previous fit, the emissivity profile is steeper
in the inner disc and flatter in the outer. With this composite model,
the 0.5--10~keV spectrum is completely reflection--dominated and only
an upper limit on the PLC normalisation can be measured (see Table~2
for details). 

\section{Spectral variability: the remaining XMM--Newton observations}

Based on our best--fit model which will be discussed below, we were
able to compute the fluxes for the six XMM--Newton
observations which are reported in Table~1 together with the exposures
obtained with the EPIC--pn camera after screening. Hereafter we define
orbits 512 and 558 as low--flux (LF) states, orbits 605, 649, and 690
as intermediate--flux (IF) states, and orbit 181 as the high--flux
(HF) state of \h\ (see Pounds et al 2004b for a slightly different
grouping of the observations). 

\begin{table}
\begin{center}
\begin{tabular}{lcccccc}
 XMM orbit &181 & 512 & 558 & 605 & 649 & 690 \\
\hline
pn exposure  &5.7 & 12.1 & 7.8 & 11.2 & 10.7 & 11.4 \\
\hline
2--10~keV Flux &15.4 & 6.5 & 6.7 & 9.6 & 8.8 & 8.5 \\
\hline
0.5--10~keV Flux &29.5& 8.2 & 9.4 & 16.7 & 14.7 & 13.7 \\
\hline
Flux state &HF& LF & LF & IF & IF & IF \\
\hline
\end{tabular}
\caption{EPIC--pn net exposures (in ks), and 2--10~keV, 0.5-10~keV
  fluxes (in $10^{-12}$~erg~s$^{-1}$~cm$^{-2}$) for the six
  XMM--Newton observations of \h. Fluxes (absorbed) are obtained from
  the best--fit models described in Section~4.1 and Table~2. Based on
  the measured flux, we divide the observations into three groups,
  corresponding to low flux (LF), intermediate flux (IF), and high
  flux (HF) states.} 
\end{center}
\end{table}

\begin{figure}
\includegraphics[width=0.30\textwidth,height=0.46\textwidth,angle=-90]{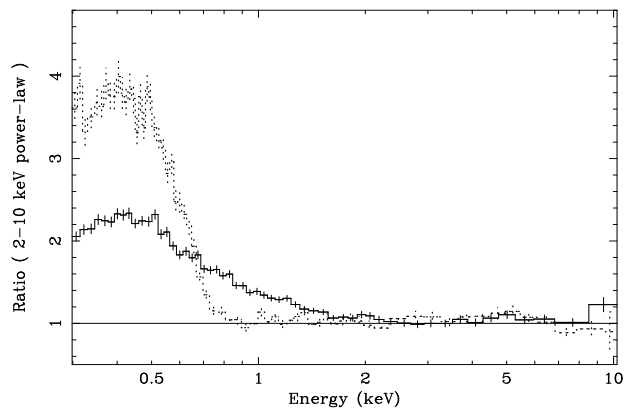}
\caption{Ratio to a power law fitted in the 2--10~keV band for a LF
  state (orbit 558, dotted grey) and a HF state (orbit 181, solid
  black) observation. The 2-10~keV photon index is flat in the LF
  state ($\Gamma =1.31$) and steeper in the HF ($\Gamma=1.80$). The
  soft excess rises steep in the LF state, while is much more gradual
  in the HF state.} 
\end{figure}

\begin{figure}
\includegraphics[width=0.34\textwidth,height=0.48\textwidth,angle=-90]{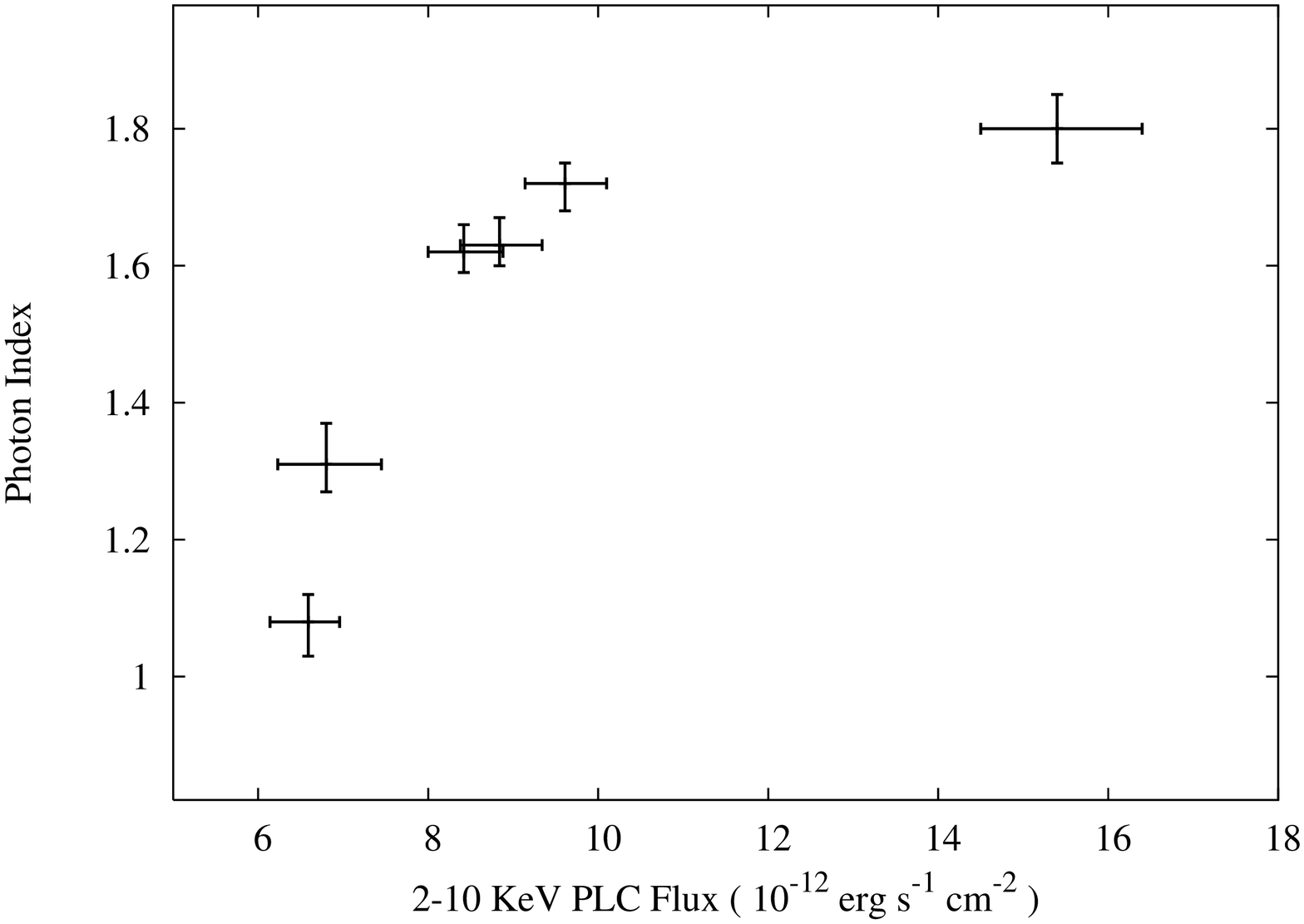}
\includegraphics[width=0.34\textwidth,height=0.48\textwidth,angle=-90]{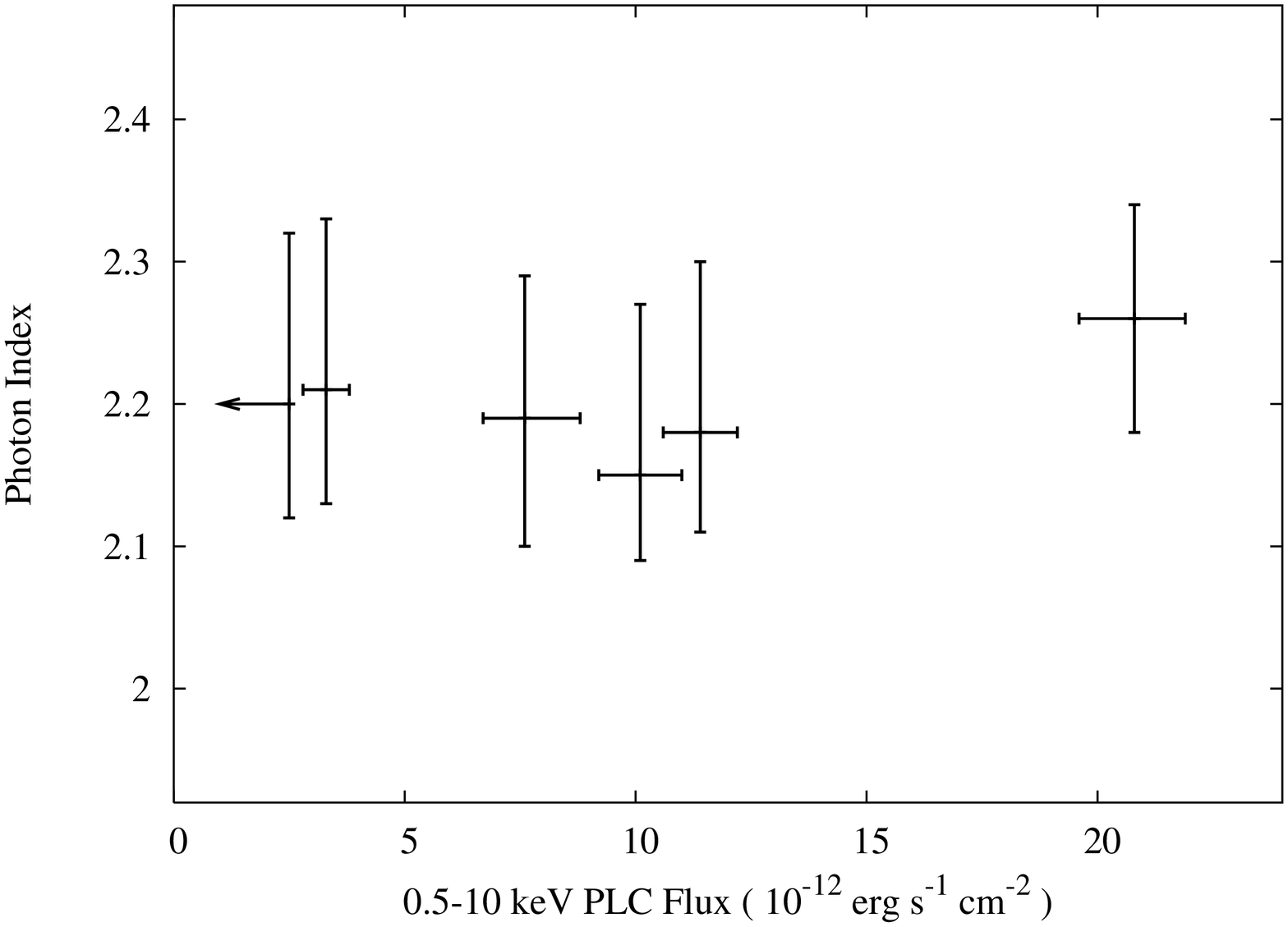}
\caption{{\bf Top}: The 2-10~keV photon index as a function of the PLC
  flux in the same band. The data have been fitted with a simple power
  law model. {\bf Bottom}: The photon index from our best--fit model
  comprising a PLC and a RDC from the accretion disc as a function of
  the PLC flux. Once the RDC is properly considered, the photon index
  appears to be consistent with a constant (with $\Gamma \simeq 2.2$).
}
\end{figure}

The XMM--Newton observations of \h\ reveal a clear trend with flux: as
shown by Pounds et al (2004b) the hard 2--10~keV spectral index
steepens, while the soft excess is more gradual as the X--ray flux
increases. As an example, in Fig.~4, we show the ratio to a power--law
model fitted between 2 and 10~keV only for two observations which are
taken as representative of a low--flux (orbit 558) and high--flux
(orbit 181) state. In the low--flux state, the 2--10~keV spectrum is
very hard with $\Gamma= 1.31 \pm 0.06$, and a steep soft excess
appears below about 0.8~keV. On the other hand, in the high--flux
state, the hard spectrum is much steeper with $\Gamma= 1.80 \pm 0.05$,
and the soft excess is already present at 1.5~keV and steepens at low
energy much more gradually.  In the top panel of Fig.~5, we plot the
2--10~keV photon index as a function of the 2--10~keV PLC flux,
showing the gradual steepening of the hard spectrum with flux. The
fitted spectral model only comprises the PLC. The same behaviour is
found in other sources (e.g. MCG--6-30-15, see Shih et al 2002) and
can often be understood in terms of a constant-$\Gamma$ PLC which
varies in normalisation affected by a weakly variable RDC. Given that
the correlation of the 2--10~keV photon index and of the soft excess
shape with flux is confirmed for all the observations, we believe that
the X-ray variability of \h\ should be explained by a model that
{\it{simultaneously}} accounts for both correlations.

\begin{table*}
\begin{center}
\begin{tabular}{lcccccc}
 XMM~orbit &181~(HF)& 512~(LF)& 558~(LF)& 605~(IF)& 649~(IF)& 690~(IF)\\
\hline
$\Gamma$  & $2.26 \pm 0.08$ &$2.20^{+0.12}_{-0.08}$&$2.21^{+0.12}_{-0.08}$&$2.18^{+0.12}_{-0.07}$&$2.15^{+0.12}_{-0.06}$&$2.19^{+0.10}_{-0.09}$\\
\\
$F_{\rm{PLC}}$  & $20.8^{+1.1}_{-1.2}$ & $<~2.5$ & $3.3\pm 0.5$ &
$11.4\pm 0.8$ & $10.1\pm 0.9$ & $7.6^{+1.2}_{-0.9}$\\
\\
$\xi_{\rm{in}}$  & $75 \pm 15$ & $69\pm 20$ & $68 \pm 18$ & $56 \pm 23$ & $56
\pm 21$ & $61\pm 21$\\
\\
$\xi_{\rm{out}}$  & $38 \pm 16$ & $29\pm 16$ & $29 \pm 13$ & $13^{+14}_{-6}$ & $12^{+18}_{-8}$ & $11^{+16}_{-4}$\\
\\
$F_{\rm{RDC}}$  & $10.1^{+2.5}_{-2.4}$ & $9.4^{+5.0}_{-2.7}$ & $7.1^{+3.3}_{-1.9}$ &
$6.1^{+1.2}_{-1.3}$ & $5.5^{+2.9}_{-1.7}$ & $6.4^{+3.7}_{-1.6}$\\
\\
$\chi^2$/dof  & 605/614 & 620/637 & 621/631 & 762/800 & 803/764 & 724/755
\end{tabular}
\caption{Results from the spectral fits to the six XMM--Newton
  observations with the composite two--component model. The 
  0.5--10~keV unabsorbed fluxes are given in units of
  $10^{-12}$~erg~cm$^{-2}$~s$^{-1}$, and the ionisation parameters in
  erg~cm~s$^{-1}$.} 
\end{center}
\end{table*}

Any constant (or weakly variable) soft emission component would
explain the spectral shape behaviour of the soft excess with flux
which would be the result of the superposition of a constant soft
emission and a variable power law component. Indeed, the soft emission
in \h\ is weakly variable as demonstrated by the analysis of the
difference spectra by Pounds et al (2004b). One remarkable result of
their analysis is that, while all the individual spectra do exhibit
soft excesses, none of the difference spectra does, showing the
constancy (or reduced variability) of the soft emission component in
\h. One possibility put forward is provided by emission from
photoionised circumnuclear gas. Given that emission from such a gas is
observed in other sources (mainly Seyfert 2 nuclei in which the gas emission
is not diluted into the primary continuum) this is certainly an
attractive scenario which however does not provide a clear and
simultaneous explanation for the 2-10~keV photon index correlation
with flux. 

An alternative can be given in terms of the two--component (RDC plus
PLC) model in which the RDC varies little while a PLC with constant
$\Gamma$ and variable normalisation determines the flux state of the
source. In such a model, the constant soft excess is due to a soft RDC
spectrum which dominates the PLC in the LF states and is more and more
contaminated by it as the flux increases, reproducing the soft
spectral shape--flux observed correlation. On the other hand, the RDC
gives rise to a very flat 2--10~keV spectrum in the LF and
reflection--dominated states, while as the flux increases and the PLC
overwhelms the RDC in the hard band, the spectrum steepens approaching
asymptotically the PLC spectral shape. This simple idea can be
visualised by considering a constant RDC and variable PLC in Fig.~3
and provides a natural, though not unique, simultaneous explanation
for the soft excess spectral shape and photon index behaviour with the
source flux. We point out here that a weakly variable RDC despite
large amplitude PLC variability is precisely the behaviour induced by
the light bending model (Miniutti \& Fabian 2004).

\subsection{Spectral analysis}

In the following, we adopt the composite--reflection model that
successfully explains the LF state of orbit 512 (Section 3) and apply
it to the remaining XMM--Newton observations. The iron abundance,
inner/outer disc radii, and disc inclination are fixed to the
best--fit values found fitting orbit 512 data. All observations are
extremely well fitted by the model, with reduced $\chi^2$ very close
to unity. The most relevant parameters and the resulting statistics
are given in Table~2 for all observations. We do not report the
parameters of the relativistic blurring which are similar to those
already reported for the LF state of orbit 512 (see Section~3), with a
very steep emissivity in the inner disc flattening in the outer
regions and a break radius around $4~r_g$. The HF state of orbit 181
exhibits a slightly flatter emissivity. In the framework of our model,
HF states correspond to situations in which the primary PLC source is
located at larger distance from the black hole than in LF states. 
Therefore, light bending is less effective, the disc illumination is
much more uniform, and the resulting emissivity is indeed expected to
be flatter. 

The first interesting result is the photon index behaviour with flux.
Once the RDC is properly taken into account, the behaviour shown in
Fig.~5 (top panel) disappears, and the measured photon index is
consistent with a constant as shown in the bottom panel of Fig.~5 and
Table~2. A value of $\Gamma \simeq 2.2$ is consistent with the
observed photon index at all flux levels, showing that the spectral
variability is not dominated by a variable spectral slope of the
primary X--ray continuum.

As pointed out by Pounds et al (2004b), a variable absorption
component local to the source is also present. We have parametrised it
with a phenomenological model comprising cold redshifted absorption
and edge roughly reproducing the presence of a warm absorber. If
our phenomenological (and rather crude) model for the warm absorber is
replaced with more sophisticated descriptions such as that provided by
the {\tt{ABSORI}} model in {\tt XSPEC} (Done et al 1992; Zdziarski et
al. 1995) with Fe abundance tied to the reflector one, no significant
differences in $\chi^2$ is seen in any of the observations. Given the
quality of the data, our phenomenological description seems sufficient
to catch the main features of the absorber and, most importantly, does
not affect our results on the RDC and PLC parameters which are our
main interest here. The edge is present in all cases with energy
around $0.74-0.78$~keV but its depth clearly diminishes with
increasing flux. The absorbing column also drops as the flux increases
from about $10^{21}$~cm$^{-2}$ to zero, showing that absorption is
more important in low than high flux states.  However, the analysis of
the difference spectra performed by Pounds et al (2004b) clearly shows
that absorption is certainly important in this source (with signatures
mainly from O, Fe, and Ne in the 0.5-1~keV band) but is not the main
driver of the variability which is instead dominated by a steep power
law component.

Our modelling allows us to identify the variable component with the
PLC providing the X--ray continuum and illuminating the accretion disc
where it is reprocessed and converted into the RDC. As mentioned,
the PLC is consistent with having constant spectral shape and only its
normalisation appears to vary. The variation of the PLC normalisation
is at least a factor 9 between the LF and HF states and determines the
flux state of \h\ (see Table~2). 

On the other hand, the RDC varies with a much smaller amplitude (about
a factor 2), and is almost consistent with being constant within the
errors at all flux levels (see Table~2). In Fig.~6, we show the
unabsorbed best--fit RDC and PLC models for a LF (orbit 558, dotted
grey) and for the HF (orbit 181, solid black) observation. The PLC
variation is about a factor 7, while the RDC is consistent with being
constant within the 90 per cent errors despite the very different flux
state. This clarifies the reason why no soft excess is measured in
difference spectra obtained between the various flux states (Pounds et
al 2004b). 

Since the RDC varies much less than the PLC, as the PLC flux drops the
spectrum becomes more and more reflection--dominated, as expected if
the light bending model applies at least qualitatively in this source.
This can be shown by plotting the reflection fraction as a function of
the PLC flux for the six observations. As a measure of the reflection
fraction, i.e. of the RDC flux relative to the PLC continuum, we
consider the $F_{\rm{RDC}}$ to $F_{\rm{PLC}}$ ratio in the whole
0.5--10~keV observable energy band. The resulting behaviour with the
PLC is shown in Fig.~7 where the anti--correlation is clearly
seen. The anti--correlation of the relative contribution of the RDC to
the total flux is one of the distinctive characteristic of the light
bending model proposed by Miniutti \& Fabian 2004 to explain the
spectral variability of X--ray sources in which a relativistic
reflection component is observed (e.g. MCG--6-30-15, 1H~0707--495, the
Galactic black hole candidate XTE~J1650--500), making \h\ one
additional candidate for observing the effects of strong gravity in
the immediate vicinity of an accreting black hole via X--ray
observations.

The observed 0.5--10~keV X--ray luminosity of the source varies from
$2\times10^{44}$~erg/s (low state) to $8.3\times 10^{44}$~erg/s (high
state), which is in the range of about 1--4 per cent of the bolometric
($2.4\times10^{46}$~erg/s, Grupe et al. 2004). In the framework of our
model, the intrinsic PLC luminosity is constant (or varies little) and
it is mostly light bending that induces the observed variations. The
intrinsic PLC luminosity can then be estimated e.g. from the
high--state (orbit 181), in which the PLC contribution to the
0.5--10~keV flux is about twice as large as the RDC (see Table~2 and
Fig.~7), and turns out to be of the order of 10--20 per cent of the
bolometric.

\begin{figure}
\includegraphics[width=0.34\textwidth,height=0.48\textwidth,angle=-90]{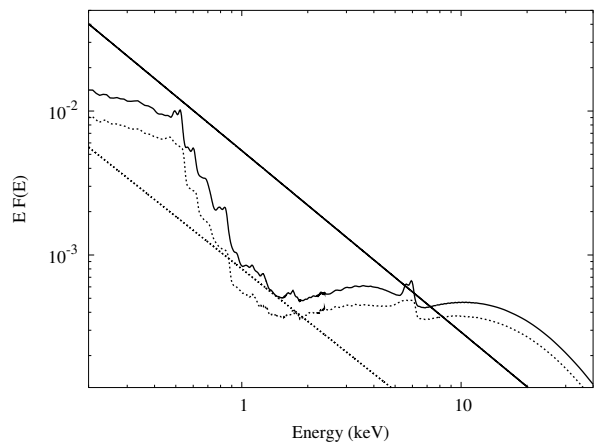}
\caption{The unabsorbed best--fit RDC and PLC models for a LF state
  observation (orbit 558, dotted grey) and a HF state one (orbit 181,
  solid black). The PLC variation is about a factor 7, while the RDC
  is consistent with being constant within the errors (see Table~2). 
  This is the reason why no (or very little) soft excess is seen in
  difference spectra between the various spectral states (Pounds et al
  2004b).  }
\end{figure}

\section{Discussion}

We have analysed all six available XMM--Newton EPIC--pn observations of \h. 
The source was observed in widely different flux states, allowing us
to group the observations in low flux (LF), intermediate flux (IF) and
high flux (HF) states and to study the spectral variability. \h\
exhibits a strong soft excess at all flux levels, but steeper in the
LF states and more gradual in the HF states. On the other hand, the
hard 2--10~keV spectrum, when modelled with a simple power law, is
clearly correlated with flux. In the LF state, the 2--10~keV photon
index is implausibly hard (about 1--1.3) for Comptonisation models,
while it seems to saturate above 1.8 in the HF state. 

We apply a two--component model to all observations comprising a PLC
and a RDC from the accretion disc, modified by absorption (both
Galactic and local to the source).  We find that the spectral
variability of \h\ can be described by a strongly variable PLC with
constant spectral shape ($\Gamma \simeq 2.2$ at all flux levels) and
an almost constant ionised RDC. The photon index is therefore found to
be much more consistent with the values allowed by standard
Comptonisation models, and the spectral variability does not appear to
be driven by photon index variations.  

The RDC is affected by strong
relativistic effects suggesting that it originates in the innermost
regions of the accretion disc around a rapidly rotating Kerr black
hole. The RDC is responsible for the soft excess and its weak
variability explains the lack of any soft excess in the difference
spectra between the various flux states, as shown in previous analysis
(Pounds et al 2004b). The difference spectra show indeed that the
variability is dominated by a steep power law which can be identified
with the PLC of our spectral decomposition. 
\begin{figure}
\includegraphics[width=0.34\textwidth,height=0.48\textwidth,angle=-90]{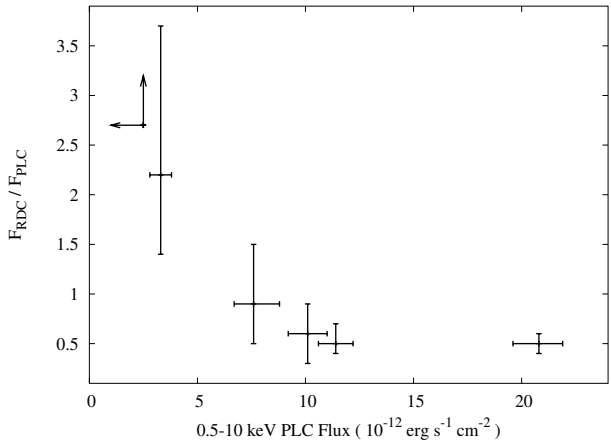}
\caption{ The ratio between the 0.5--10~keV flux of the RDC to the PLC
  is shown as a function of the PLC flux. This is a measure of the
  reflection fraction, i.e., the relative contribution of the RDC to
  the total flux. The observed anti--correlation is in very good
  agreement with the predictions of the light bending model (Miniutti
  \& Fabian 2004). 
} 
\end{figure}

A major consequence of the above variability is that the spectrum
becomes more and more reflection--dominated as the flux drops. Indeed,
our analysis reveals that the spectrum of the lowest flux observation
(orbit 512) can be interpreted as almost completely
reflection--dominated. The RDC simultaneously explains both the soft
excess, the flat 2--10~keV spectral shape, and the broad residuals in
the Fe K band during the LF state. The parameters of the relativistic
blurring and the behaviour of the reflection fraction with flux match
the prediction of our light bending model, in which the primary source
of the PLC is located in the region of strong gravity within a few
gravitational radii from the central black hole. 

Given the quality of the present data, the model we are proposing is
not unique and other interpretations of the X--ray spectrum and
variability of \h\ are possible. Pounds et al (2004a; 2004b) consider
as preferential a model in which the soft emission comprises an almost
constant ``core'' component due to emission from photoionised gas,
while the soft excess is entirely an artifact of absorption by a warm
absorber. We point out that observing soft X--ray emission by
photoionised gas would be somewhat unusual in such a high luminosity
source. On the other hand, the flat photon index
of the low flux state is explained by Pounds et al with the presence
of a substantial column of additional cold matter that covers only
partially the source and produces an absorption edge and spectral
curvature in the Fe K band. 

Our spectral decomposition has the great advantage of explaining the
broadband spectrum and variability with the interplay of only two
(broadband) components without the need of invoking separate
components in the different energy bands. In addition it provides a
natural explanation for the remarkably uniform temperature of
100--200~eV found in the soft X--ray spectra of a large number of low
redshift active galaxies with soft excess (Gierlinski \& Done 2004,
Porquet et al 2004) despite large differences in black hole mass and
luminosity (hence in accretion disc temperatures). The very narrow
range of temperatures might indicate an origin in atomic (rather than
truly thermal) processes for the soft excess. Absorption, as proposed
by Pounds et al (2004b) for \h, provides a possibility, but some
active galaxies exhibit a soft excess despite the lack of any
absorption signature (see e.g. Ark~120, Vaughan et al 2004).  The
other natural explanation involves relativistically blurred ionised
reflection (and bremsstrahlung from the disc surface layers). The
ionised reflection model we are using is indeed well fitted by a
100--200~eV blackbody for a wide range of ionisation and blurring
parameters, when folded through the EPIC--pn response matrix in the
0.3-2~keV band (Ross \& Fabian 2005). 

We note that the small inner disc radius required by our model, with
$r_{\rm{in}} \le 2~r_g$ in the lowest flux state, and the relatively
low ionisation of the disc material, indicate that high density gas
extends down to those radii.  Although magnetic field and mass
accretion rate nonaxisymmetric variations can distort the inner
reflecting radius of a disc (Krolik \& Hawley 2002), we doubt that the
innermost stable circular orbit (ISCO) can be much further out than
$2~r_g$. Since the ISCO is related to the spin parameter $a$ of the
black hole by a simple formula (see e.g. Bardeen, Press \& Teukolsky
1972), the upper limit on $r_{\rm{in}}$ translates into a lower limit
for the black hole spin parameter in \h\ of $a \ge 0.95$. 

A much longer XMM--Newton exposure of \h\ during a low flux state
would help resolving the partial covering/broad iron line degeneracy. 
Moreover, the forthcoming launch of the Astro--E2 X--ray satellite,
with unprecedented energy resolution in the Fe K band, could help
resolving this issue in \h\ and other sources. If the edge detected at
7.1~keV is due to absorption, it will be reasonably sharper than if it
is due to the combination of a reflection edge and the blue wing of a
relativistic line, the difference between the two models being within
the grasp of the Astro--E2 capabilities. The 2--10~keV luminosity of
\h\ makes this source a potential important target for Astro--E2, as
opposed to 1H~0707-495 which is about one order of magnitude fainter
in the 5--7~keV band. 

\section*{Acknowledgments}
Based on observations obtained with XMM-Newton, an ESA science mission
with instruments and contributions directly funded by ESA Member
States and NASA. We also thank Ken Pounds for kindly providing the
data that were used in this work. ACF thanks the Royal
Society for support. GM and KI thank the PPARC, and RRR the College of
the Holy Cross for support.

\end{document}